\documentclass[conference]{IEEEtran}
\usepackage{amsfonts}
\IEEEoverridecommandlockouts

\ifCLASSINFOpdf
\else
\fi
\usepackage{epsfig}
\usepackage{graphicx}
\usepackage{subfigure}
\usepackage{psfig}
\usepackage{epsf}
\usepackage[cmex10]{amsmath}
\usepackage{booktabs}
\usepackage{fancyhdr}
\newtheorem{myTheo}{Theorem}   

\newtheorem{rem}{Remark} 
\begin{document}
\title{Robust Max-Min Fairness Energy Harvesting in Secure MISO Cognitive Radio With SWIPT}
\author{Fuhui Zhou$^\dagger$, Zan Li$^\dagger$, Julian Cheng$^\ddag$, Qunwei Li$^\S $, and Jiangbo Si$^\dagger$\\
$^\dagger$ State Key Laboratory of Integrated Services Networks,
Xidian University, Xi$^{'}$an, China\\
$^\ddag$ The University of British Columbia, Kelowna, BC, Canada\\
$^\S $ The Department of Electrical Engineering and Computer Science, Syracuse University, Syracuse, USA.\\
Email: \emph{\{fuhuizhou@stu.xidian.edu.cn, \{zanli, jbsi\}@xidian.edu.cn, julian.cheng@ubc.ca, qli33@syr.edu\}}
\thanks{The research reported in this article was supported in part by the National Natural Science Foundation of China (61301179, 61501356, 61501354, 61401323 and 61401338).}}
\maketitle
\begin{abstract}
A multiple-input single-output cognitive radio downlink network is studied with simultaneous wireless information and power transfer. In this network, a secondary user coexists with multiple primary users and multiple energy harvesting receivers. In order to guarantee secure communication and energy harvesting, the problem of robust secure artificial noise-aided beamforming and power splitting design is investigated under imperfect channel state information. Specifically, the max-min fairness energy harvesting problem is formulated under the bounded channel state information error model. A one-dimensional search algorithm based on ${\cal S}\text{-Procedure} $ is proposed to solve the problem. It is shown that the optimal robust secure beamforming can be achieved. A tradeoff is elucidated between the secrecy rate of the secondary user receiver and the energy harvested by the energy harvesting receivers under a max-min fairness criterion.
\end{abstract}
\IEEEpeerreviewmaketitle
\section{Introduction}
\IEEEPARstart{C}{OGNITIVE} radio (CR) is a promising technique that aims to alleviate the spectrum scarcity problem \cite{Haykin}. In CR under spectrum sharing, a secondary user (SU) can coexist with a primary user (PU) as long as the interference caused by the SU is tolerable to the PU. Besides the spectrum scarcity problem, the escalating requirement for higher data rate and the unprecedented increase of mobile devices have contributed to the sharp growth of energy consumption and resulted in energy scarcity problem, especially for CR with energy-limited devices, such as energy-limited wireless sensors and cellular phones \cite{C. Jiang}-\cite{X. Lu}. In order to address the energy scarcity problem, a number
of works focused on maximizing the energy efficiency of CR networks \cite{Y. Pei11}, \cite{F. H. Zhou}.

Recently, a promising technology called simultaneous wireless information and power transfer (SWIPT) has been proposed to address the energy scarcity problem, and it has attracted much research attention \cite{X. Lu}. Specifically, radio frequency (RF) signals not only can carry information, but also can be used as a source for wireless power transfer, which can charge the energy-limited communication devices. Thus, in CR with energy-limited communication devices, it is of great importance to investigate CR with SWIPT that can improve spectrum efficiency and energy utilization simultaneously \cite{D. W. K. Ng1}, \cite{B. Fang}. However, due to the inherent characteristics of CR with SWIPT, malicious energy harvesting receivers (EHRs) may exist and illegitimately access the PU bands or change the radio environment. As a result, the legitimate SU is unable to use frequency bands of the PU or has his confidential transmitted information intercepted \cite{D. W. K. Ng1}-\cite{C. Xu}. Thus, the security of CR with SWIPT is also of crucial importance.

Physical-layer security, which is based on the physical layer characteristics of the wireless channels, has been proposed to improve the security of wireless communication systems \cite{D. W. K. Ng1}. However, it was shown that the secrecy rate of a wireless communication system with physical-layer security is limited by the channel state information (CSI). In CR, the secrecy rate of the SU is further limited since the transmit power of the SU should be controlled to protect the PU from harmful interference \cite{Y. Pei}. In order to improve the secrecy rate of the SU, both multi-antenna technique and beamforming technique have been introduced \cite{Y. Pei}-\cite{C. Wang}. However, the optimal beamforming schemes proposed in \cite{Y. Pei}-\cite{C. Wang} may not be appropriate in CR with SWIPT since energy harvested by EHRs should be considered. On the other hand, in practice, it is difficult to obtain the perfect CSI due to the existence of channel estimation errors and quantization errors, especially in CR with SWIPT, where there is no cooperation among SUs, PUs and EHRs. Even worse, the imperfect CSI can significantly deteriorate the beamforming performance. Thus, it is of great importance to design robust secure beamforming for CR with SWIPT.

In secure CR with SWIPT, the robust design of beamforming not only provides the SU a reasonably high secrecy rate, but also protects the PU from harmful interference. Moreover, in secure CR with SWIPT, the robust beamforming design can guarantee the energy harvesting requirement and prolong the operation time. Although many investigations into the robust beamforming design issue in the traditional wireless communication systems with SWIPT were reported in \cite{D. W. K. Ng2}-\cite{F. Wang}, few investigations have been devoted to designing robust secure beamforming schemes for CR with SWIPT. Moreover, the fairness among different users needs to be considered in the design of future wireless communication systems, but it was neglected in previous studies \cite{D. W. K. Ng1}-\cite{F. Wang}.

In this paper, the energy harvested by EHRs under a max-min fairness criterion is optimized by jointly designing robust artificial noise (AN)-aided beamforming and the power splitting ratio in MISO CR with SWIPT. The formulated problem is non-convex and challenging. In order to solve the problem, a one-dimensional search algorithm is proposed based on the semidefinite relaxation (SDR) and ${\cal S}$-Procedure \cite{S. P. Boyd}. It is proved that the optimal robust secure beamforming can always be found and the rank of the optimal AN covariance is one. A tradeoff is found between the secrecy rate of the SU and the energy harvested by EHRs under a max-min fairness criterion.

The rest of this paper is organized as follows. Section II presents the system model. The max-min fairness energy harvesting problem is formulated in Section III. Section IV presents simulation results. Section V concludes the paper.

\emph{Notations:} Matrices and vectors are denoted by boldface capital letters and boldface lower case letters. $\mathbf{I}$ denotes the identity matrix. The Hermitian (conjugate) transpose, trace, and rank of a matrix \textbf{A} is denoted by $\mathbf{A^H}$, Tr$\left(\mathbf{A}\right)$ and $\text{r}\left(\mathbf{A}\right)$. $\mathbf{x}^\dag$ represents the conjugate transpose of a vector $\mathbf{x}$. $\mathbf{C}^{M\times N}$ stands for a $M$-by-$N$ dimensional complex matrix set. $\mathbf{A}\succeq \mathbf{0} \left(\mathbf{A}\succ \mathbf{0}\right)$ represents that $\mathbf{A}$ is a Hermitian positive semidefinite (definite) matrix. $\mathbb{H}^N$ and $\mathbb{H}_+^{N}$ represent a $N$-by-$N$ dimensional Hermitian matrix set and a Hermitian positive semidefinite matrix set. ${\left| \cdot \right|}$ represents the absolute value of a complex scalar. $\mathbf{x} \sim {\cal C}{\cal N}\left( {\mathbf{u},\mathbf{\Sigma } }\right)$ means that $\mathbf{x}$ is a random vector, which follows a complex Gaussian distribution with mean $\mathbf{u}$ and covariance $\mathbf{\Sigma }$. $\mathbb{E}[ \cdot ]$ denotes the expectation operator. $\mathbb{R}_{+}$ represents the set of all nonnegative real numbers.
\section{System Model}
A downlink MISO CR network with SWIPT under spectrum sharing is considered, where one secondary link, $M$ primary links, and $K$ energy harvesting links share the same spectrum. The cognitive base station (CBS) is equipped with $N_t$ antennas. The primary base station (PBS), the SU receiver, each PU receiver and each EHR are equipped with single antenna. The PBS transmits information to $M$ PU receivers while the CBS provides a SWIPT service to the SU receiver, and transfers energy to $K$ EHRs as long as the interference imposed on $M$ PU receivers from the CBS is tolerable. It is assumed that the SU receiver is an energy-limited device such as a cellular phone and a wireless sensor whose battery has limited energy storage capacity \cite{X. Huang}-\cite{B. Fang}. Different from \cite{D. W. K. Ng1}, a power splitting receiver architecture is adopted here in the SU receiver. Thus, the SU receiver splits the received signal power into two parts in order to simultaneously decode information and harvest energy. Due to the inherent characteristics of CR and SWIPT, EHRs may eavesdrop and intercept the information transmitted by the CBS. In this paper, we assume that PU receivers are friendly users and do not eavesdrop the information sent by the CBS. This assumption has been widely used in secure CR \cite{Y. Pei}-\cite{C. Wang}. All the channels involved are assumed to be slow frequency-nonselective fading channels. Let ${\cal K}$ and ${\cal I}$ denote the set ${\cal K} \buildrel \Delta \over = \left\{ {1,2, \cdots ,K} \right\}$ and the set ${\cal I} \buildrel \Delta \over = \left\{ {1,2, \cdots ,M} \right\}$, respectively. The signals received at the SU receiver and the $k$th EHR, and the interference signals imposed on the $i$th PU receiver from the CBS, denoted by $y$, ${y_{e,k}}$, and ${y_{PU,i}}$, where $k \in \cal K$ and $i \in \cal I$, can be given, respectively, as
\begin{subequations}
\begin{align}\label{27}\
&y = {\mathbf{h}^\dag }\mathbf{x} + {n_s}\\
&{y_{e,k}} = \mathbf{g}_k^\dag \mathbf{x} + {n_{e,k}},\ k \in \cal K\\
&{y_{PU,i}} = \mathbf{q}_i^\dag \mathbf{x},\ i \in \cal I
\end{align}
\end{subequations}
where $\mathbf{h} \in {\mathbf{C}^{{N_t} \times 1}}$, $\mathbf{g}_k \in {\mathbf{C}^{{N_t} \times 1}}$ and $\mathbf{q}_k \in {\mathbf{C}^{{N_t} \times 1}}$ are the channel vectors between the CBS and the SU receiver, the $k$th EHR and the $i$th PU receiver, respectively. In $\left(1\right)$, ${n_s} \sim {\cal C}{\cal N}\left( {0,\sigma _s^2} \right)$ and ${n_{e,k}} \sim {\cal C}{\cal N}\left( {0,\sigma _e^2} \right)$ denote complex Gaussian noise at the SU receiver and the $k$th EHR, respectively, which include the interference from the PBS and the additive complex Gaussian noise at the SU receiver and the $k$th EHR. Note that the interferences at the SU receiver and the $k$th EHR from the PBS are assumed to be circularly symmetric complex Gaussian. The model for the interference is the worst-case model and has been widely used in \cite{Y. Pei11}-\cite{D. W. K. Ng1}. This model also covers the system models studied in \cite{B. Fang}-\cite{C. Wang}, where the interferences from the PBS to the SU receiver and the $k$th EHR are negligible. $\mathbf{x} \in {\mathbf{C}^{{N_t} \times 1}}$ is the transmit signal vector, given as
\begin{align}\label{27}\
\mathbf{x} = \mathbf{w}s + \mathbf{v}
\end{align}
where $s \in {\mathbf{C}^{{1} \times 1}}$ and $\mathbf{w}\in {\mathbf{C}^{{N_t} \times 1}}$ denote the confidential information-bearing signal for the SU receiver and the corresponding beamforming vector, respectively. Without loss of generality, it is assumed that $\mathbb{E}[ {{{\left| s \right|}^2}} ] = 1$. In $\left(2\right)$, $\mathbf{v}\in {\mathbf{C}^{{N_t} \times 1}}$ is the noise vector artificially generated by the CBS in order to improve the secrecy rate of the SU and energy transfer at EHRs. It is assumed that $\mathbf{v} \sim {\cal C}{\cal N}\left( {0,\mathbf{\Sigma} } \right)$, where $\mathbf{\Sigma}$ is the AN covariance to be designed. Since the SU receiver is equipped with a power splitting device, the equivalent signal received at the SU receiver for information decoding, denoted by ${y_D}$, is given as
\begin{align}\label{27}\
{y_D} = \sqrt \rho  \left( {{ \mathbf{h}^\dag } \mathbf{x} + {n_s}} \right) + {n_{s,p}}
\end{align}
where ${n_{s,p}}$ is the additional processing noise generated by the SU receiver. It is assumed that ${n_{s,p}} \sim {\cal C}{\cal N}\left( {0,\sigma _{s,p}^2} \right)$ and ${n_{s,p}}$ is independent of ${n_s}$ and ${n_{e,k}}$. $\rho  \in \left( {0,1} \right)$ denotes the power splitting ratio, which denotes that $\rho$ portion of the received power is used to decode information and $1-\rho$ portion of the received power is utilized to energy transfer. Thus, the secrecy rate of the SU receiver, denoted by $R_s$, is given as
\begin{subequations}
\begin{align}\label{27}\
&{R_s} ={\mathop {\min }\limits_{k \in \cal K} }\ {\left\{ { C_s- C_{e,k}} \right\}}\\
& C_s={{\log }_2}\left( {1 + \frac{{\rho {\mathbf{h}^\dag}\mathbf{w}{\mathbf{w}^\dag}\mathbf{h}}}{{\rho \left( {{\mathbf{h}^\dag}\mathbf{\Sigma} \mathbf{h} + \sigma _s^2} \right) + \sigma _{s,p}^2}}} \right)\\
& C_{e,k}={{\log }_2}\left( {1 + \frac{{\mathbf{g}_k^\dag\mathbf{w}{\mathbf{w}^\dag}{\mathbf{g}_k}}}{{\mathbf{g}_k^\dag\mathbf{\Sigma} {\mathbf{g}_k} + \sigma _e^2}}} \right)
\end{align}
\end{subequations}
and the energy harvested at the SU receiver and the $k$th EHR, denoted by $E_s$ and $E_{e,k}$, can be respectively expressed as
\begin{subequations}
\begin{align}\label{27}\
&{E_s} = \left( {1 - \rho } \right)\eta \left( {{\mathbf{h}^\dag }\mathbf{w}{\mathbf{w}^\dag }\mathbf{h} + {\mathbf{h}^\dag }\mathbf{\Sigma} \mathbf{h} + \sigma _s^2} \right)\\
&{E_{e,k}} = \eta \left( {\mathbf{g}_k^\dag \mathbf{w}{\mathbf{w}^\dag }{\mathbf{g}_k} + \mathbf{g}_k^\dag \mathbf{\Sigma} {\mathbf{g}_k} + \sigma _e^2} \right),\ k \in \cal K
\end{align}
\end{subequations}
where $\eta  \in \left( {0,\left. 1 \right]} \right.$ is a constant, denoting the energy conversion efficiency at the SU receiver and the EHRs. It is assumed that the CSI $\mathbf{h}$ is available at both CBS and SU \cite{Y. Pei11}-\cite{Y. Pei}. However, the channel vectors $\mathbf{g}_k, k \in \cal K$ and $\mathbf{q}_i, i \in \cal I$ cannot be perfectly known since there is no cooperation among the CBS, PU receivers and EHRs \cite{D. W. K. Ng1}, \cite{Y. Pei1}-\cite{D. W. K. Ng2}. In this paper, the bounded CSI error model is applied, which uses a bounded set to model the CSI errors \cite{D. W. K. Ng1}, \cite{C. Xu}-\cite{F. Wang}. In this case, the worst-case performance can be guaranteed and the interference from the CBS to PUs can be strictly controlled under any realization of CSI errors in the bounded set. Moreover, CR with the bounded CSI error model can be implemented with low complexity. The bounded CSI error model for the channel vector $\mathbf{g}_k, k \in \cal K$, is given as
\begin{subequations}
\begin{align}\label{27}\
&{\mathbf{g}_k} = \overline {{\mathbf{g}_k}}  + \Delta {\mathbf{g}_k},\ k \in \cal K\\
&{\mathbf{\Psi} _{e,k}} \buildrel \Delta \over = \left\{ {\Delta {\mathbf{g}_k} \in {\mathbf{C}^{{N_t} \times 1}}:\Delta \mathbf{g}_k^\dag \Delta {\mathbf{g}_k} \le \xi _{e,k}^2} \right\}
\end{align}
\end{subequations}
and the bounded CSI error model for the channel vector $\mathbf{q}_i, i \in \cal I$, is given as
\begin{subequations}
\begin{align}\label{27}\
&{\mathbf{q}_i} = \overline {{\mathbf{q}_i}}  + \Delta {\mathbf{q}_i},\ i \in \cal I\\
&{\mathbf{\Psi} _{P,i}} \buildrel \Delta \over = \left\{ {\Delta {\mathbf{q}_i} \in {\mathbf{C}^{{N_t} \times 1}}:\Delta \mathbf{q}_i^\dag \Delta {\mathbf{q}_i} \le \xi _{P,i}^2} \right\}
\end{align}
\end{subequations}
where $\overline {{\mathbf{g}_k}}$ and $\overline {{\mathbf{q}_i}}$ are the estimates of the channel vectors $\mathbf{g}_k$ and $\mathbf{q}_m$, respectively; ${\mathbf{\Psi} _{e,k}} $ and ${\mathbf{\Psi} _{P,i}}$ denote the uncertainty regions of $\mathbf{g}_k$ and $\mathbf{q}_m$; $\Delta {\mathbf{g}_k}$ and $\Delta {\mathbf{q}_i}$ represent the channel estimation errors of $\mathbf{g}_k$ and $\mathbf{q}_m$; $\xi _{e,k}$ and $\xi _{P,i}$ are the radiuses of the uncertainty regions, ${\mathbf{\Psi} _{e,k}} $ and ${\mathbf{\Psi} _{P,i}}$, respectively.
\section{Max-Min Fairness Energy Harvesting Problem}
Let $\mathbf{W}$ and $\mathbf{H}$ denote $\mathbf{W} = \mathbf{w}{\mathbf{w}^\dag }$ and $\mathbf{H} = \mathbf{h}{\mathbf{h}^\dag }$. Based on the bounded CSI error model, the max-min fairness energy harvesting problem in MISO CR with SWIPT, denoted by $\textbf{P}_{{1}}$, is given as
\begin{subequations}
\begin{align}\label{27}\
&\textbf{P}_{{1}}: \ {\mathop {\max }\limits_{\mathbf{W},\mathbf{\Sigma} } } \ {\mathop {\min }\limits_{\Delta {\mathbf{g}_k} \in {\mathbf{\Psi }_{e,k}}, k \in \cal{K}} } {\eta \left( {\mathbf{g}_k^\dag\left( {\mathbf{W} + \mathbf{\Sigma} } \right){\mathbf{g}_k} + \sigma _e^2} \right)}\\
&\text{s.t.}\  C1:{R_s} \ge R_{\min},\ \forall \Delta {\mathbf{g}_k} \in {\mathbf{\Psi} _{e,k}},\ k \in \cal K\\
&\  C2:\left( {1 - \rho } \right)\eta \left( {\text{Tr}\left( {\mathbf{W}\mathbf{H} + \mathbf{\Sigma} \mathbf{H}} \right) + \sigma _s^2} \right) \ge {\psi _s}\\
& \  C3:\mathbf{q}_i^\dag \left( {\mathbf{W} + \mathbf{\Sigma} } \right){\mathbf{q}_i} \le {P_{In,i}},\ \forall \Delta {\mathbf{q}_i} \in {\mathbf{\Psi} _{P,i}},\ i \in \cal I\\
& \  C4:\text{Tr}\left( {\mathbf{W} + \mathbf{\Sigma} } \right) \le {P_{th}}\\
& \  C5:0 < \rho  < 1\\
&  \  C6:\text{r}\left( \mathbf{W} \right) = 1\\
& \  C7:\mathbf{\Sigma} \succeq0, \ \mathbf{W}\succeq0
\end{align}
\end{subequations}
where $R_{\min}$ is the minimum secrecy rate requirement of the CBS; ${\psi _s}$ denotes the minimum EH required at the SU receiver; ${P_{In,i}}$ represents the maximum tolerable interference power of the $i$th PU receiver; $P_{th}$ is the maximum transmit power of the CBS. In $\left(8\rm{b}\right)$-$\left(8\rm{i}\right)$, $C1$ can guarantee that the secrecy rate of the SU is not less than $R_{\min}$; $C2$ is the EH constrains of the SU receiver; $C3$ is imposed in order to protect the QoS of PUs; $C4$ limits the maximum transmit power of the CBS. Note that $\textbf{P}_{{1}}$ may be infeasible due to the transmit power constraint $C4$. Owing to the non-convex constraints in $C2$ and $C6$, $\textbf{P}_{{1}}$ is non-convex and difficult to be solved. Moreover, there are infinite inequality constraints to be satisfied due to the uncertain regions, ${\mathbf{\Psi} _{e,k}} $ and ${\mathbf{\Psi} _{P,i}}$, which make $\textbf{P}_{{1}}$ even more challenging. In order to solve $\textbf{P}_{{1}}$, a slack variable $\tau\geq0$ is introduced. Using the slack variable, $\tau$, $\textbf{P}_{{1}}$ can be equivalently expressed as the following problem, denoted by $\textbf{P}_{{2}}$, given as
\begin{subequations}
\begin{align}\label{27}\
& \textbf{P}_{{2}}: \ {\mathop {\max }\limits_{\mathbf{W},\mathbf{\Sigma} } }\  \tau \\
&\text{s.t.}\ C1-C7 \\
& \eta \left( {\mathbf{g}_k^\dag \left( {\mathbf{W} + \mathbf{\Sigma} } \right){\mathbf{g}_k} + \sigma _e^2} \right) \ge \tau ,\Delta {\mathbf{g}_k} \in {\mathbf{\Psi } _{e,k}},k \in \cal K.
\end{align}
\end{subequations}
In order to solve $\textbf{P}_{{2}}$, SDR and ${\cal S}$-Procedure are applied.

\emph{\textbf{Lemma 1 $\left({\cal S}\text{-Procedure} \right) $ }} \cite{S. P. Boyd}: Let $ {f_i}\left( \mathbf{z} \right) = {\mathbf{z}^\dag }{\mathbf{A}_i}\mathbf{z} + 2{\mathop{\mathbf{\rm Re}}\nolimits} \left\{ {\mathbf{b}_i^\dag \mathbf{z}} \right\} + {c_i}, i \in \left\{ {1,2} \right\}$, where $\mathbf{z}\in \mathbf{C}^{N\times 1}$, $\mathbf{A}_i \in \mathbb{H}^N$, $\mathbf{b}_i\in \mathbf{C}^{N\times 1}$ and $c_i\in \mathbb{R}$. Then, the expression ${f_1}\left( \mathbf{z} \right) \le 0 \Rightarrow {f_2}\left( \mathbf{z} \right) \le 0$ holds if and only if there exists a $\alpha  \ge 0$ such that
\begin{align}\label{27}\
\alpha \left[ {\begin{array}{*{20}{c}}
{{\mathbf{A}_1}}&{{\mathbf{b}_1}}\\
{\mathbf{b}_1^\dag }&{{c_1}}
\end{array}} \right] - \left[ {\begin{array}{*{20}{c}}
{{\mathbf{A}_2}}&{{\mathbf{b}_2}}\\
{\mathbf{b}_2^\dag }&{{c_2}}
\end{array}} \right]
\succeq \mathbf{0}
\end{align}
provided that there exists a vector $\mathbf{\widehat z}$ such that ${f_i}\left( {\mathbf{\widehat z}} \right) < 0$.

By introducing slack variables, $t=1/\rho$, and $\beta$, where $t>1$ and $\beta\geq1$, and applying \textbf{Lemma 1}, the constraints $C1$, $C2$, $C3$ and the constraint given by $\left(9\rm{c}\right)$, can be equivalently expressed as $\left(11\right)$ at the top of the next page. In $\left(11\right)$, ${\omega _k} \ge 0, k\in \cal K$, $\mu _k\geq0$ and $\delta _i\geq0, i\in \cal I$ are slack variables. Note that $\left(11\rm{a}\right)$ and $\left(11\rm{b}\right)$ are obtained from $C1$.
\begin{figure*}[!t]
\normalsize
\begin{subequations}
\begin{align}\label{27}\
&\mathbf{Tr}\left\{ {\left( {\mathbf{W} + \left( {1 - {2^{{R_{\min }}}}\beta } \right)\mathbf{\Sigma} } \right)\mathbf{H}} \right\} + \left( {1 - {2^{{R_{\min }}}}\beta } \right)\left( {\sigma _s^2 + \sigma _{s,p}^2t} \right) \ge 0\\ \notag
&{\mathbf{\Gamma} _k}\left( {{\omega _k},\mathbf{W},\mathbf{\Sigma} ,\beta } \right)\\ &=\left[ {\begin{array}{*{20}{c}}
{{\omega _k}\mathbf{I} - \left( {\mathbf{W} - \left( {\beta  - 1} \right)\mathbf{\Sigma} } \right)}&{ - \left( {\mathbf{W} - \left( {\beta  - 1} \right)\mathbf{\Sigma} } \right)\overline {{\mathbf{g}_k}} }\\
{ - {{\overline {{\mathbf{g}_k}} }^\dag }\left( {\mathbf{W} - \left( {\beta  - 1} \right)\mathbf{\Sigma} } \right)}&{\left( {\beta  - 1} \right)\sigma _e^2 - {{\overline {{\mathbf{g}_k}} }^\dag }\left( {W - \left( {\beta  - 1} \right)\mathbf{\Sigma} } \right)\overline {{\mathbf{g}_k}}  - {\omega _k}\xi _{e,k}^2}
\end{array}} \right]\succeq\mathbf{0} ,\ k\in \cal K\\
&\mathbf{Tr}\left\{ {\left( {\mathbf{W} + \mathbf{\Sigma} } \right)\mathbf{H}} \right\} + \sigma _s^2 - \frac{{{\psi _s}}}{\eta }\left( {1 + \frac{1}{{t - 1}}} \right) \ge 0\\
&{\mathbf{\Gamma} _i}\left( {{\delta _i},\mathbf{W},\mathbf{\Sigma} } \right)=\left[ {\begin{array}{*{20}{c}}
{{\delta _i}\mathbf{I} - \left( {\mathbf{W} + \mathbf{\Sigma} } \right)}&{ - \left( {\mathbf{W} + \mathbf{\Sigma} } \right)\overline {{\mathbf{q}_i}} }\\
{ - {{\overline {{\mathbf{q}_i}} }^\dag }\left( {\mathbf{W} + \mathbf{\Sigma} } \right)}&{{P_{In,i}} - {{\overline {{\mathbf{q}_i}} }^\dag }\left( {\mathbf{W} + \mathbf{\Sigma} } \right)\overline {{\mathbf{q}_i}}  - {\delta _i}\xi _{P,i}^2}
\end{array}} \right]\succeq\mathbf{0},\ \ i\in \cal I\\
&{\mathbf{\Gamma} _k}\left( {{\mu _k},\mathbf{W},\mathbf{\Sigma}, \tau } \right)=\left[ {\begin{array}{*{20}{c}}
{{\mu _k}\mathbf{I} + \left( {\mathbf{W} + \mathbf{\Sigma} } \right)}&{\left( {\mathbf{W} + \mathbf{\Sigma} } \right)\Delta {\mathbf{g}_k}}\\
{{{\overline {{\mathbf{g}_k}} }^\dag }\left( {\mathbf{W} + \mathbf{\Sigma} } \right)}&{{{\overline {{\mathbf{g}_k}} }^\dag }\left( {\mathbf{W} + \mathbf{\Sigma} } \right)\overline {{\mathbf{g}_k}}  + \sigma _e^2 - \tau{\eta ^{ - 1}} - {\mu _k}\xi _{e,k}^2}
\end{array}} \right]\succeq\mathbf{0}, \ \ k\in \cal K
\end{align}
\end{subequations}
\hrulefill \vspace*{4pt}
\end{figure*}
By using SDR, $\textbf{P}_{{2}}$ can be relaxed to the following problem, given as
\begin{subequations}
\begin{align}\label{27}\
\textbf{P}_{{3}}: \ \ \ \ \ &{\mathop {\max }\limits_{\mathbf{W},\mathbf{\Sigma} ,t,\beta, \{\omega _k\},\{\mu _k\},\{\delta _i\} } }\  \tau \\
&\text{s.t.}\ \ \  t  > 1\\
&\ \ \ \ \ \ \  \beta\geq1\\
&\ \ \ \ \ \ \  \omega _k, \mu _k, \delta _i\geq0\\
&\ \ \ \ \ \   \left(11\right), C4 \ \text{and}\ C7.
\end{align}
\end{subequations}
Unfortunately, since the variable $\beta$ couples the other variables, $\textbf{P}_{{3}}$ is still non-convex. Using the fact
that $R_{\min}\geq0$, $\left(8\rm{b}\right)$ and $\left(8\rm{e}\right)$, on can obtain $\beta  \le 1 + {P_{th}}{{\left\| \mathbf{h} \right\|}^2}/\sigma _{s,p}^2$. Thus, the following inequalities can be obtained
\begin{align}\label{27}\
1 \le \beta  \le 1 + \frac{{{P_{th}}}}{{\sigma _{s,p}^2}}{\left\| \mathbf{h} \right\|^2}.
\end{align}
It is seen from $\left(12\right)$ that $\textbf{P}_{{2}}$ is convex for a given $\beta$. Thus, the optimal $\beta$ value can be searched from the interval $[ {1,1 + {P_{th}}{{\left\| \mathbf{h} \right\|}^2}/\sigma _{s,p}^2}]$. Using the uniform sampling method to search the optimal $\beta$, \textbf{Algorithm 1} is proposed to solve $\textbf{P}_{{2}}$. Table I summarizes the details of \textbf{Algorithm 1}. We can now state the following two theorems.
\begin{table}[htbp]
\begin{center}
\caption{The one-dimensional line search algorithm}
\begin{tabular}{lcl}
\\\toprule
$\textbf{Algorithm 1}$: The one-dimensional line search algorithm\\ \midrule
\  1: \textbf{Setting:}\\
\ \ \ \ \ \ \ $R_{\min}$, ${\psi _s}$ and ${\psi _{e,k}},\ k \in \cal K$, ${P_{In,i}},\ i\in\cal I$, $P_{th}$, $\xi _{e,k}$, $\xi _{P,i}$\\
\ \ \ \ \ \ \ and the uniform search step, $\tau_1$.\\
\  2: \textbf{Initialization:}\\
\ \ \ \ \ \ \ the iteration index $n=1$. \\
\  3: \textbf{Optimization:}\\
\ 4:\ \ \ \ \textbf{$\pmb{\unrhd} $  for  $\beta$=1:$\tau_1$:$1 + {P_{th}}{{\left\| \mathbf{h} \right\|}^2}/\sigma _{s,p}^2$}\\
\ 5:\ \ \ \ \ \ \ \  \ use the software \texttt{CVX} to solve $\textbf{P}_{{4}}$ for the given $\beta$, given as \\
\ \ \ \ \ \ \ \ \ \ \ \  \ $ \textbf{P}_{{4}}: \ \ \ \ \ {\mathop {\min }\limits_{\mathbf{W},\mathbf{\Sigma} ,t, \{\omega _k\},\{\mu _k\},\{\delta _i\} } }\ {\tau} $\\
\ \ \ \ \ \ \ \ \ \ \ \ \ \ \ \ \ \ \ \  $ \text{s.t.} \ \ \ \left(12\rm{b}\right), \left(12\rm{d}\right)\ \text{and} \ \left(12\rm{e}\right).$ \\
\ 6:\ \ \ \ \ \ \ \ obtain $\mathbf{W}^n$, $\mathbf{\Sigma}^n$, $t^n$, and $ {\tau^n  } $;\\
\ 7: \ \ \ \ \ \ \ set $n=n+1$;\\
\ 8:\ \ \ \ \textbf{$\pmb{\unrhd} $ end} \\
\  9: \textbf{Optimal design:}\\
\ 10:\ \ \ \ \  \  ${\tau _{opt}} = \arg {\mathop {\max }\limits_{} } \ \tau^n$, $\mathbf{W}^{opt}=\left[\mathbf{W}^n\right]_{\tau  = {\tau _{opt}}}$ \\
\ \ \ \ \  \ \ \ \ \ $\mathbf{\Sigma}^{opt}=\left[\mathbf{\Sigma}^n\right]_{\tau  = {\tau _{opt}}}$ and $\rho^{opt}=\left[1/t^{n}\right]_{\tau  = {\tau _{opt}}}$.\\
\bottomrule
\end{tabular}
\end{center}
\end{table}

\begin{myTheo}
For $\textbf{P}_{{3}}$, assume that the minimum secrecy rate $R_{\min}>0$ and that $\textbf{P}_{{3}}$ is feasible, the optimal $\mathbf{W}$ of $\textbf{P}_{{3}}$ is unique and its rank is one.
\end{myTheo}
\begin{IEEEproof}
See Appendix A.
\end{IEEEproof}
\begin{rem}
Theorem 1 indicates that the max-min fairness energy harvesting problem given as $\textbf{P}_{{1}}$ can be solved by solving the rank-relaxed problem presented in $\textbf{P}_{{3}}$. The optimal transmit matrix $\mathbf{W}$ is rank-one in spite of the numbers of the PU receivers and EHRs. Moreover, for any $\beta$, the optimal $\mathbf{W}$ is rank-one. Thus, the optimal robust secure beamforming vector can be obtained by performing eigenvalue decomposition over $\mathbf{W}$, and it is the eigenvector related to the maximum eigenvalue of $\mathbf{W}$.
\end{rem}
\begin{myTheo}
The optimal AN covariance of $\textbf{P}_{{3}}$, $\mathbf{\Sigma}^{opt}$, has the rank-one property regardless of the number of PU receivers and the number of the EHRs.
\end{myTheo}

\begin{IEEEproof}
The proof is similar to the proof for Theorem 1 and it is omitted due to space limitation.
\end{IEEEproof}

\section{Simulation Results}
In this section, simulation results are given to illustrate the performance of the proposed robust secure beamforming scheme. The simulation results to be presented are based on the following simulation settings. The numbers of PU receivers and EHRs are set as $M=2$ and $N=3$, respectively. The energy conservation efficiency, $\eta$, is $1$. The maximum tolerable interference power levels of all PU receivers are set as $P_{In,i}=-10$ dB, $i\in \mathcal{I}$. The maximum transmit power of the CBS is $P_{th}=2$ dB. The variances of noise at the SU receiver, the EH receivers, and the variance of the processing noise at the SU receiver are $\sigma _s^2=0.1$, $\sigma _e^2=0.1$ and $\sigma _{s,p}^2=0.01$, respectively. The minimum EH of the SU receiver is set as ${\psi _s}=22$ dBm. All the involved channels are assumed to be Rayleigh flat fading, i.e., $ {\mathbf{h}} \sim {\cal C}{\cal N}\left( {0,{\mathbf{I}}} \right)$, $\overline {{\mathbf{g}_k}} \sim {\cal C}{\cal N}\left( {0,{\mathbf{I}}} \right), \ k \in \cal K$, and $\overline {{\mathbf{q}_i}} \sim {\cal C}{\cal N}\left( {0,{0.1\mathbf{I}}} \right), \ i \in \cal I$. $\xi _{e,k}^2$ for all $k$ and $\xi _{P,i}^2$ for all $i$ are set as $\xi _{e,k}^2=0.001$ and $\xi _{P,i}^2=0.0001$, respectively.
\begin{figure}[h]
    \begin{minipage}[b]{0.45\textwidth}
      \centering
      \includegraphics[height=6cm,width=8cm]{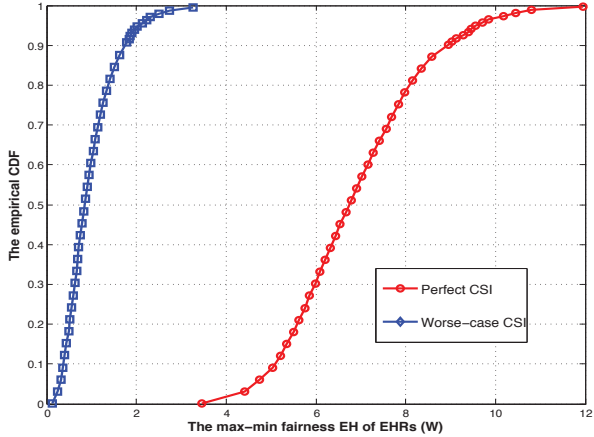}
      \vskip-0.2cm\centering {\footnotesize (a)}
    \end{minipage}\hfill
    \begin{minipage}[b]{0.45\textwidth}
      \centering
      \includegraphics[height=6cm,width=8cm]{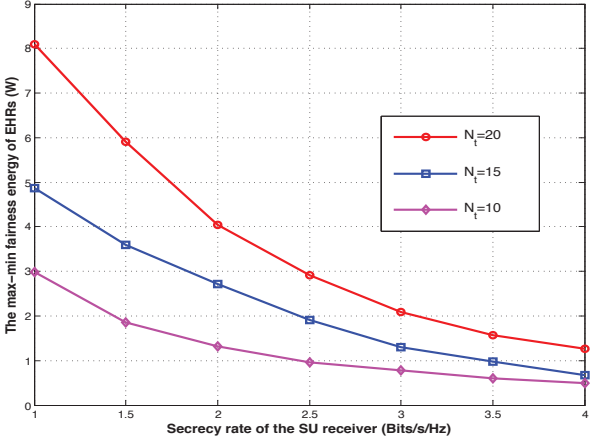}
      \vskip-0.2cm\centering {\footnotesize (b)}
    \end{minipage}\hfill
    \begin{minipage}[b]{0.45\textwidth}
      \centering
      \includegraphics[height=6cm,width=8cm]{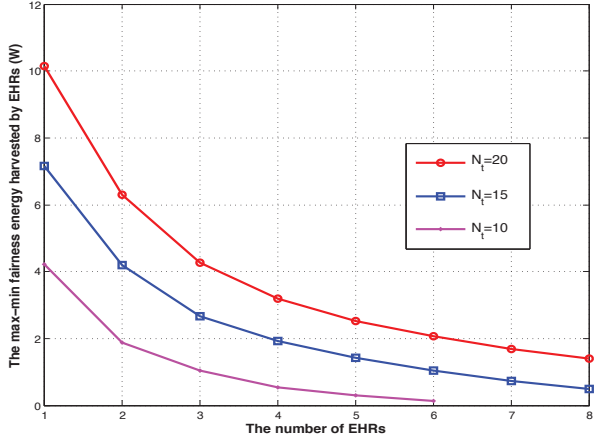}
      \vskip-0.2cm\centering {\footnotesize (b)}
    \end{minipage}\hfill
 \caption{(a)~The empirical CDF of the max-min fairness energy harvested by EHRs, $R_{\min}=1.5$ bits/s/Hz; (b)~The max-min fairness energy harvested by EHRs versus the secrecy rate of the SU receiver; (c)~The max-min fairness energy harvested by EHRs versus the number of EHRs.}
  \label{fig: Scenario_2} 
\end{figure}

Figure 1(a) shows the empirical cumulative distribution functions (CDFs) of the max-min fairness energy harvested by EHRs under the bounded CSI error model and under the perfect CSI. The minimum secrecy rate of the SU receiver is set as $R_{\min}=1.5$ bits/s/Hz. The empirical CDFs shown in Fig. 1(a) are obtained by using 1,000 channel realizations. As shown in Fig. 1(a), the max-min fairness energy of EHRs obtained under the perfect CSI is larger than that attained under the bounded CSI error model. This is due to the fact that the CBS requires more transmit power to guarantee that the SU receiver obtains the minimum secrecy rate when the CSI is imperfect.

The max-min fairness energy harvested by EHRs versus the secrecy rate of the SU receiver under the bounded CSI error model is presented in Fig. 1(b). The number of the transmit antennas of the CBS is set as $N_t=10, 15,\ \text{or} \ 20$. It is seen from Fig. 1(b) that the max-min fairness energy of EHRs decreases with the increase of the minimum secrecy rate requirement of the SU receiver. This indicates that there is a tradeoff between the energy harvested by EHRs under a max-min fairness criterion and the secrecy rate of the SU receiver. As shown in Fig. 1(b), the max-min fairness energy harvested by EHRs increases with the number of the transmit antennas of the CBS. It is explained by the fact that the diversity gain increases when the number of the transmit antennas of the CBS is increased.

Figure 1(c) shows the max-min fairness energy harvested by EHRs versus the number of EHRs under the bounded CSI error model. The minimum secrecy rate of the SU receiver is set as $R_{\min}=2$ bits/s/Hz. The number of the transmit antennas of the CBS is $N_t=10, 15,\ \text{or} \ 20$. It is seen from Fig. 1(c) that the max-min fairness energy harvested by EHRs decreases with the increase of the number of EHRs. It can be explained by the fact that the robust secure beamforming is required to be designed to guarantee the secrecy rate of the SU receiver, which is susceptible to the number of the EHRs. It is observed that the max-min fairness EH problem is infeasible when the number of EHRs is larger than $6$ and $N_t=10$. The reason is that the minimum secrecy rate of the SU receiver can not be satisfied due to the maximum transmit power constraint. This is consistent with our analysis presented in Section III.
\section{Conclusion}
Secure communication was studied for a MISO CR downlink network with SWIPT. Robust secure beamforming and power splitting ratio were jointly designed under the bounded CSI error model in order to guarantee secure communication and energy harvested by EHRs. The energy harvested by EHRs was optimized under a max-min fairness criterion. A one-dimensional search algorithm was proposed to solve the challenging non-convex problem. It was proved that the optimal robust secure beamforming vector and the optimal AN covariance matrix can always be found. A tradeoff was found between the secure rate of the SU receiver and the energy harvested by EHRs under a max-min fairness criterion.
\appendices
\section{Proof of Theorem 1}
The proof for Theorem 1 is based on the Karush-Kuhn-Tucker optimality conditions of $\textbf{P}_{{3}}$. Let $\mathbf{\Xi}$ denote a collection of all the dual and primal variables related to $\textbf{P}_{{3}}$. Then, the Lagrangian of problem $\textbf{P}_{{3}}$ can be given by $\left(14\right)$ at the top of the next page.
\begin{figure*}[!t]
\normalsize
\begin{align}\label{27}\ \notag
\cal {L} \left(  \mathbf{\Xi} \right) =& \tau + \alpha \left\{ {\text{Tr}\left\{ {\left( {\mathbf{W} + \left( {1 - {2^{{R_{\min }}}}\beta } \right)\mathbf{\Sigma} } \right)\mathbf{H}} \right\} + \left( {1 - {2^{{R_{\min }}}}\beta } \right)\left( {\sigma _s^2 + t\sigma _{s,p}^2} \right)} \right\}\\ \notag
& + \sum\limits_{k = 1}^K {\text{Tr}\left\{ {{\mathbf{A}_k}{\mathbf{\Gamma} _k}\left( {{\omega _k},\mathbf{W},\mathbf{\Sigma} ,\beta } \right)} \right\}}  + \sum\limits_{k = 1}^K {\text{Tr}\left\{ {{\mathbf{B}_k}{\mathbf{\Gamma} _k}\left( {{\mu _k},\mathbf{W},\mathbf{\Sigma}, \tau } \right)} \right\}}  + \sum\limits_{i = 1}^M {\text{Tr}\left\{ {{\mathbf{D}_i}{\mathbf{\Gamma} _i}\left( {{\delta _i},\mathbf{W},\mathbf{\Sigma} } \right)} \right\}} \\ \notag
& + {\nu _1}\left\{ {\left( {\text{Tr}\left( {\mathbf{W}\mathbf{H} + \mathbf{\Sigma}\mathbf{H}} \right) + \sigma _s^2} \right) - \left( {1 + \frac{1}{{t - 1}}} \right)\frac{1}{\eta }{\psi _s}} \right\} - {\nu _2}\left( {\text{Tr}\left( {\mathbf{W} + \mathbf{\Sigma} } \right) - {P_{th}}} \right) \\
 &+ \text{Tr}\left( {\mathbf{W}\mathbf{Y}} \right) + \text{Tr}\left( {\mathbf{\Sigma} \mathbf{Z}} \right)+ \Lambda
\end{align}
\hrulefill \vspace*{4pt}
\end{figure*}
$ \alpha\in \mathbb{R}_{+}$, $ \nu _1\in \mathbb{R}_{+}$ and $ \nu _2\in \mathbb{R}_{+}$, are the dual variables with respect to $\left(11\rm{a}\right)$, $\left(11\rm{c}\right)$ and $C4$, respectively. $\Lambda$ denotes the collection of terms involving the variables, which are not related to the proof. ${\mathbf{A}_k}\in \mathbb{H}_+^{N}$, ${\mathbf{B}_k}\in \mathbb{H}_+^{N}$, ${\mathbf{D}_i}\in \mathbb{H}_+^{N}$, $\mathbf{Y}\in \mathbb{H}_+^{N}$ and $\mathbf{Z}\in \mathbb{H}_+^{N}$ are the dual variables with respect to $\left(11\rm{b}\right)$, $\left(11\rm{e}\right)$, $\left(11\rm{d}\right)$ and $\left(8\rm{h}\right)$, respectively. Let ${ \mathbf{\Lambda} _k} = [ {\begin{array}{*{20}{c}}
 \mathbf{I}&{\overline {{ \mathbf{g}_k}} }
\end{array}} ]
$ and ${\mathbf{\Upsilon} _i} = [ {\begin{array}{*{20}{c}}
\mathbf{I}&{\overline {{\mathbf{q}_i}} }
\end{array}} ]
$. Then, ${\mathbf{\Gamma} _k}\left( {{\omega _k},\mathbf{W},\mathbf{\Sigma} ,\beta } \right)$, ${\mathbf{\Gamma} _k}\left( {{\mu _k},\mathbf{W},\mathbf{\Sigma}, \tau } \right)$ and ${\mathbf{\Gamma} _i}\left( {{\delta _i},\mathbf{W},\mathbf{\Sigma} } \right)$ can be rewritten as
\begin{subequations}
\begin{align}\label{27}\
&{\mathbf{\Gamma} _k}\left( {{\omega _k},\mathbf{W},\mathbf{\Sigma} ,\beta } \right) = \left[ {\begin{array}{*{20}{c}}
{{\omega _k}\mathbf{I}}&0\\ \notag
0&{\left( {\beta  - 1} \right)\sigma _e^2 - {\omega _k}\xi _{e,k}^2}
\end{array}} \right]\\
 &\ \ \ \ \ \ \ \ \ \ \ \ \ \ \ \ \ \ \ \ \ \ \ \ - \mathbf{\Lambda} _k^H\left( {\mathbf{W}- \left( {\beta  - 1} \right)\mathbf{\Sigma} } \right){\mathbf{\Lambda} _k},
\\ \notag
& {\mathbf{\Gamma} _k}\left( {{\mu _k},\mathbf{W},\mathbf{\Sigma},\tau } \right)= \left[ {\begin{array}{*{20}{c}}
{{\mu _k}\mathbf{I}}&0\\
0&{\sigma _e^2 - {\tau}{\eta ^{ - 1}} - {\mu _k}\xi _{e,k}^2}
\end{array}} \right] \\
&\ \ \ \ \ \ \ \ \ \ \ \ \ \ \ \ \ \ \ \ \ \ \ \ + \mathbf{\Lambda} _k^H\left( {\mathbf{W} + \mathbf{\Sigma} } \right){\mathbf{\Lambda} _k}, k\in \cal K,\\ \notag
&{\mathbf{\Gamma} _i}\left( {{\delta _i},\mathbf{W},\mathbf{\Sigma} } \right) = \left[ {\begin{array}{*{20}{c}}
{{\delta _i}\mathbf{I}}&0\\
0&{{P_{In,i}} - {\delta _i}\xi _{P,i}^2}
\end{array}} \right] \\
&\ \ \ \ \ \ \ \ \ \ \ \ \ \ \ \ \ \ \ - \mathbf{\Upsilon} _i^H\left( {\mathbf{W} + \mathbf{\Sigma} } \right){\mathbf{\Upsilon} _i}, i\in \cal I.
\end{align}
\end{subequations}
The partial KKT conditions related to the proof can be given as
\begin{subequations}
\begin{align}\label{27}\ \notag
& \alpha \mathbf{H} + \sum\limits_{k = 1}^K {{\mathbf{\Lambda} _k}\left( {{\mathbf{B}_k} - {\mathbf{A}_k}} \right)\mathbf{\Lambda}_k^H}  - \sum\limits_{i = 1}^M {\mathbf{\Upsilon} _i{\mathbf{D}_i}\mathbf{\Upsilon} _i^H}  + {\nu _1}\mathbf{H} \\
&- {\nu _2}\mathbf{I} + \mathbf{Y} = \mathbf{0},
\\
& \mathbf{Y}\mathbf{W} = \mathbf{0},\\
&t = \frac{{\sqrt {\eta \alpha \left( { {2^{{R_{\min }}}}\beta }-1 \right)} {\sigma _{s,p}} + \sqrt {{\nu _1}{\psi _s}} }}{{\sqrt {\eta \alpha \left( {{2^{{R_{\min }}}}\beta } -1\right)} {\sigma _{s,p}}}},\\
&{\mathbf{A}_k},{\mathbf{B}_k},{\mathbf{D}_i}\succeq\mathbf{0}, \alpha, {\nu _1},{\nu _2} \ge 0.
\end{align}
\end{subequations}
Since $t>1$, it can be obtained from $\left(16\rm{c}\right)$ that $\alpha>0$ and $\nu _1>0$. Right-multiplying $\left(16\rm{a}\right)$ by $\mathbf{W}$ and combining $\left(16\rm{b}\right)$, one has
\begin{align}\label{27}\ \notag
&\left( { { {\nu _2}} \mathbf{I} + \sum\limits_{i = 1}^M {{\mathbf{\Upsilon} _i}{\mathbf{D}_i}\mathbf{\Upsilon} _i^H}  + \sum\limits_{k = 1}^K {{\mathbf{\Lambda}_k}\left( {{\mathbf{A}_k} - {\mathbf{B}_k}} \right)\mathbf{\Lambda} _k^H} } \right)\mathbf{W}\\
& = \left( {\alpha  + {\nu _1}} \right)\mathbf{H}\mathbf{W}.
\end{align}
According to $\left(17\right)$, one has
\begin{subequations}
\begin{align}\label{27}\ \notag
&\text{r}\left\{\left( { {{\nu _2}} \mathbf{I} + \sum\limits_{i = 1}^M {{\mathbf{\Upsilon} _i}{\mathbf{D}_i}\mathbf{\Upsilon} _i^H}  + \sum\limits_{k = 1}^K {{\mathbf{\Lambda}_k}\left( {{\mathbf{A}_k} - {\mathbf{B}_k}} \right)\mathbf{\Lambda} _k^H} } \right)\mathbf{W} \right\}\\
&= \text{r}\left\{\left( {\alpha  + {\nu _1}} \right)\mathbf{H}\mathbf{W}\right\},
\\
& \text{r}\left\{\left( {\alpha  + {\nu _1}} \right)\mathbf{H}\mathbf{W}\right\}\leq1.
\end{align}
\end{subequations}
According to $\left(16\rm{a}\right)$, one has
\begin{align}\label{27}\ \notag
&\left( { { {\nu _2}} \mathbf{I} + \sum\limits_{i = 1}^M {{\mathbf{\Upsilon} _i}{\mathbf{D}_i}\mathbf{\Upsilon} _i^H}  + \sum\limits_{k = 1}^K {{\mathbf{\Lambda}_k}\left( {{\mathbf{A}_k} - {\mathbf{B}_k}} \right)\mathbf{\Lambda} _k^H} } \right) \\
&= \mathbf{Y}+\left( {\alpha  + {\nu _1}} \right)\mathbf{H}.
\end{align}
Since $\mathbf{Y}+\left( {\alpha  + {\nu _1}} \right)\mathbf{H}\succ \mathbf{0}$,  one obtains the following relationship
\begin{align}\label{27}\ \notag
&\text{r}\left(\mathbf{W}\right)\\ \notag
&=\text{r}\left\{\left( { { {\nu _2}} \mathbf{I} + \sum\limits_{i = 1}^M {{\mathbf{\Upsilon} _i}{\mathbf{D}_i}\mathbf{\Upsilon} _i^H}  + \sum\limits_{k = 1}^K {{\mathbf{\Lambda}_k}\left( {{\mathbf{A}_k} - {\mathbf{B}_k}} \right)\mathbf{\Lambda} _k^H} } \right)\mathbf{W} \right\}\\
&= \text{r}\left\{\left( {\alpha  + {\nu _1}} \right)\mathbf{H}\mathbf{W}\right\}\leq1.
\end{align}
Thus, if $\textbf{P}_{{3}}$ is feasible and $R_{\min}>0$, the rank of $\mathbf{W}$ is one. The proof is completed.

\end{document}